
%
\catcode`@=11 
%
%
%

\font\fourteenrm=cmr10 scaled\magstep2
\font\twelverm=cmr10 scaled\magstep1
\font\ninerm=cmr9            \font\sixrm=cmr6

\font\fourteenbf=cmbx10 scaled\magstep2
\font\twelvebf=cmbx10 scaled\magstep1
\font\ninebf=cmbx9            \font\sixbf=cmbx6
\font\seventeeni=cmmi10 scaled\magstep3     \skewchar\seventeeni='177
\font\fourteeni=cmmi10 scaled\magstep2      \skewchar\fourteeni='177
\font\twelvei=cmmi10 scaled\magstep1        \skewchar\twelvei='177
\font\ninei=cmmi9                           \skewchar\ninei='177
\font\sixi=cmmi6                            \skewchar\sixi='177
\font\seventeensy=cmsy10 scaled\magstep3    \skewchar\seventeensy='60
\font\fourteensy=cmsy10 scaled\magstep2     \skewchar\fourteensy='60
\font\twelvesy=cmsy10 scaled\magstep1       \skewchar\twelvesy='60
\font\ninesy=cmsy9                          \skewchar\ninesy='60
\font\sixsy=cmsy6                           \skewchar\sixsy='60

\font\fourteenex=cmex10 scaled\magstep2
\font\twelveex=cmex10 scaled\magstep1

\font\fourteensl=cmsl10 scaled\magstep2
\font\twelvesl=cmsl10 scaled\magstep1
\font\ninesl=cmsl9

\font\fourteenit=cmti10 scaled\magstep2
\font\twelveit=cmti10 scaled\magstep1
\font\twelvett=cmtt10 scaled\magstep1
\font\twelvecp=cmcsc10 scaled\magstep1
\font\tencp=cmcsc10
\newfam\cpfam
%
%
\newcount\f@ntkey            \f@ntkey=0
\def\samef@nt{\relax \ifcase\f@ntkey \rm \or\oldstyle \or\or
         \or\it \or\sl \or\bf \or\tt \or\caps \fi }
\def\fourteenpoint{\relax
    \textfont0=\fourteenrm          \scriptfont0=\tenrm
    \scriptscriptfont0=\sevenrm
     \def\rm{\fam0 \fourteenrm \f@ntkey=0 }\relax
    \textfont1=\fourteeni           \scriptfont1=\teni
    \scriptscriptfont1=\seveni
     \def\oldstyle{\fam1 \fourteeni\f@ntkey=1 }\relax
    \textfont2=\fourteensy          \scriptfont2=\tensy
    \scriptscriptfont2=\sevensy
    \textfont3=\fourteenex     \scriptfont3=\fourteenex
    \scriptscriptfont3=\fourteenex
    \def\it{\fam\itfam \fourteenit\f@ntkey=4 }\textfont\itfam=\fourteenit
    \def\sl{\fam\slfam \fourteensl\f@ntkey=5 }\textfont\slfam=\fourteensl
    \scriptfont\slfam=\tensl
    \def\bf{\fam\bffam \fourteenbf\f@ntkey=6 }\textfont\bffam=\fourteenbf
    \scriptfont\bffam=\tenbf     \scriptscriptfont\bffam=\sevenbf
    \def\tt{\fam\ttfam \twelvett \f@ntkey=7 }\textfont\ttfam=\twelvett
    \h@big=11.9\p@{} \h@Big=16.1\p@{} \h@bigg=20.3\p@{} \h@Bigg=24.5\p@{}
    \def\caps{\fam\cpfam \twelvecp \f@ntkey=8 }\textfont\cpfam=\twelvecp
    \setbox\strutbox=\hbox{\vrule height 12pt depth 5pt width\z@}
    \samef@nt}
\def\twelvepoint{\relax
    \textfont0=\twelverm          \scriptfont0=\ninerm
    \scriptscriptfont0=\sixrm
     \def\rm{\fam0 \twelverm \f@ntkey=0 }\relax
    \textfont1=\twelvei           \scriptfont1=\ninei
    \scriptscriptfont1=\sixi
     \def\oldstyle{\fam1 \twelvei\f@ntkey=1 }\relax
    \textfont2=\twelvesy          \scriptfont2=\ninesy
    \scriptscriptfont2=\sixsy
    \textfont3=\twelveex          \scriptfont3=\twelveex
    \scriptscriptfont3=\twelveex
    \def\it{\fam\itfam \twelveit \f@ntkey=4 }\textfont\itfam=\twelveit
    \def\sl{\fam\slfam \twelvesl \f@ntkey=5 }\textfont\slfam=\twelvesl
    \scriptfont\slfam=\ninesl
    \def\bf{\fam\bffam \twelvebf \f@ntkey=6 }\textfont\bffam=\twelvebf
    \scriptfont\bffam=\ninebf     \scriptscriptfont\bffam=\sixbf
    \def\tt{\fam\ttfam \twelvett \f@ntkey=7 }\textfont\ttfam=\twelvett
    \h@big=10.2\p@{}
    \h@Big=13.8\p@{}
    \h@bigg=17.4\p@{}
    \h@Bigg=21.0\p@{}
    \def\caps{\fam\cpfam \twelvecp \f@ntkey=8 }\textfont\cpfam=\twelvecp
    \setbox\strutbox=\hbox{\vrule height 10pt depth 4pt width\z@}
    \samef@nt}
\def\tenpoint{\relax
    \textfont0=\tenrm          \scriptfont0=\sevenrm
    \scriptscriptfont0=\fiverm
    \def\rm{\fam0 \tenrm \f@ntkey=0 }\relax
    \textfont1=\teni           \scriptfont1=\seveni
    \scriptscriptfont1=\fivei
    \def\oldstyle{\fam1 \teni \f@ntkey=1 }\relax
    \textfont2=\tensy          \scriptfont2=\sevensy
    \scriptscriptfont2=\fivesy
    \textfont3=\tenex          \scriptfont3=\tenex
    \scriptscriptfont3=\tenex
    \def\it{\fam\itfam \tenit \f@ntkey=4 }\textfont\itfam=\tenit
    \def\sl{\fam\slfam \tensl \f@ntkey=5 }\textfont\slfam=\tensl
    \def\bf{\fam\bffam \tenbf \f@ntkey=6 }\textfont\bffam=\tenbf
    \scriptfont\bffam=\sevenbf     \scriptscriptfont\bffam=\fivebf
    \def\tt{\fam\ttfam \tentt \f@ntkey=7 }\textfont\ttfam=\tentt
    \def\caps{\fam\cpfam \tencp \f@ntkey=8 }\textfont\cpfam=\tencp
    \setbox\strutbox=\hbox{\vrule height 8.5pt depth 3.5pt width\z@}
    \samef@nt}
%
%
%
%
\newdimen\h@big  \h@big=8.5\p@
\newdimen\h@Big  \h@Big=11.5\p@
\newdimen\h@bigg  \h@bigg=14.5\p@
\newdimen\h@Bigg  \h@Bigg=17.5\p@
\def\big#1{{\hbox{$\left#1\vbox to\h@big{}\right.\n@space$}}}
\def\Big#1{{\hbox{$\left#1\vbox to\h@Big{}\right.\n@space$}}}
\def\bigg#1{{\hbox{$\left#1\vbox to\h@bigg{}\right.\n@space$}}}
\def\Bigg#1{{\hbox{$\left#1\vbox to\h@Bigg{}\right.\n@space$}}}
%
%
%
\normalbaselineskip = 20pt plus 0.2pt minus 0.1pt
\normallineskip = 1.5pt plus 0.1pt minus 0.1pt
\normallineskiplimit = 1.5pt
\newskip\normaldisplayskip
\normaldisplayskip = 20pt plus 5pt minus 10pt
\newskip\normaldispshortskip
\normaldispshortskip = 6pt plus 5pt
\newskip\normalparskip
\normalparskip = 6pt plus 2pt minus 1pt
\newskip\skipregister
\skipregister = 5pt plus 2pt minus 1.5pt
\newif\ifsingl@    \newif\ifdoubl@
\newif\iftwelv@    \twelv@true
\def\singlespace{\singl@true\doubl@false\spaces@t}
\def\doublespace{\singl@false\doubl@true\spaces@t}
\def\normalspace{\singl@false\doubl@false\spaces@t}
\def\Tenpoint{\tenpoint\twelv@false\spaces@t}
\def\Twelvepoint{\twelvepoint\twelv@true\spaces@t}
\def\spaces@t{\relax%
 \iftwelv@ \ifsingl@\subspaces@t3:4;\else\subspaces@t29:31;\fi%
 \else \ifsingl@\subspaces@t3:5;\else\subspaces@t4:5;\fi \fi%
 \ifdoubl@ \multiply\baselineskip by 5%
 \divide\baselineskip by 4 \fi \unskip}
\def\subspaces@t#1:#2;{
      \baselineskip = \normalbaselineskip
      \multiply\baselineskip by #1 \divide\baselineskip by #2
      \lineskip = \normallineskip
      \multiply\lineskip by #1 \divide\lineskip by #2
      \lineskiplimit = \normallineskiplimit
      \multiply\lineskiplimit by #1 \divide\lineskiplimit by #2
      \parskip = \normalparskip
      \multiply\parskip by #1 \divide\parskip by #2
      \abovedisplayskip = \normaldisplayskip
      \multiply\abovedisplayskip by #1 \divide\abovedisplayskip by #2
      \belowdisplayskip = \abovedisplayskip
      \abovedisplayshortskip = \normaldispshortskip
      \multiply\abovedisplayshortskip by #1
        \divide\abovedisplayshortskip by #2
      \belowdisplayshortskip = \abovedisplayshortskip
      \advance\belowdisplayshortskip by \belowdisplayskip
      \divide\belowdisplayshortskip by 2
      \smallskipamount = \skipregister
      \multiply\smallskipamount by #1 \divide\smallskipamount by #2
      \medskipamount = \smallskipamount \multiply\medskipamount by 2
      \bigskipamount = \smallskipamount \multiply\bigskipamount by 4 }
\def\normalbaselines{ \baselineskip=\normalbaselineskip
   \lineskip=\normallineskip \lineskiplimit=\normallineskip
   \iftwelv@\else \multiply\baselineskip by 4 \divide\baselineskip by 5
     \multiply\lineskiplimit by 4 \divide\lineskiplimit by 5
     \multiply\lineskip by 4 \divide\lineskip by 5 \fi }
\Twelvepoint  
\interlinepenalty=50
\interfootnotelinepenalty=5000
\predisplaypenalty=9000
\postdisplaypenalty=500
\hfuzz=1pt
\vfuzz=0.2pt
%
%
%
\def\pagecontents{
   \ifvoid\topins\else\unvbox\topins\vskip\skip\topins\fi
   \dimen@ = \dp255 \unvbox255
   \ifvoid\footins\else\vskip\skip\footins\footrule\unvbox\footins\fi
   \ifr@ggedbottom \kern-\dimen@ \vfil \fi }
\def\makeheadline{\vbox to 0pt{ \skip@=\topskip
      \advance\skip@ by -12pt \advance\skip@ by -2\normalbaselineskip
      \vskip\skip@ \line{\vbox to 12pt{}\the\headline} \vss
      }\nointerlineskip}
\def\makefootline{\baselineskip = 1.5\normalbaselineskip
                 \line{\the\footline}}
\newif\iffrontpage
\newif\ifletterstyle
\newif\ifp@genum
\def\nopagenumbers{\p@genumfalse}
\def\pagenumbers{\p@genumtrue}
\pagenumbers
\newtoks\paperheadline
\newtoks\letterheadline
\newtoks\letterfrontheadline
\newtoks\lettermainheadline
\newtoks\paperfootline
\newtoks\letterfootline
\newtoks\date
\footline={\ifletterstyle\the\letterfootline\else\the\paperfootline\fi}
\paperfootline={\hss\iffrontpage\else\ifp@genum\tenrm\folio\hss\fi\fi}
\letterfootline={\hfil}
\headline={\ifletterstyle\the\letterheadline\else\the\paperheadline\fi}
\paperheadline={\hfil}
\letterheadline{\iffrontpage\the\letterfrontheadline
     \else\the\lettermainheadline\fi}
\lettermainheadline={\rm\ifp@genum page \ \folio\fi\hfil\the\date}
\def\monthname{\relax\ifcase\month 0/\or January\or February\or
   March\or April\or May\or June\or July\or August\or September\or
   October\or November\or December\else\number\month/\fi}
\date={\monthname\ \number\day, \number\year}
\countdef\pagenumber=1  \pagenumber=1
\def\advancepageno{\global\advance\pageno by 1
   \ifnum\pagenumber<0 \global\advance\pagenumber by -1
    \else\global\advance\pagenumber by 1 \fi \global\frontpagefalse }
\def\folio{\ifnum\pagenumber<0 \romannumeral-\pagenumber
           \else \number\pagenumber \fi }
\def\footrule{\dimen@=\prevdepth\nointerlineskip
   \vbox to 0pt{\vskip -0.25\baselineskip \hrule width 0.35\hsize \vss}
   \prevdepth=\dimen@ }
\newtoks\foottokens
\foottokens={\Tenpoint\singlespace}
\newdimen\footindent
\footindent=24pt
\def\vfootnote#1{\insert\footins\bgroup  \the\foottokens
   \interlinepenalty=\interfootnotelinepenalty \floatingpenalty=20000
   \splittopskip=\ht\strutbox \boxmaxdepth=\dp\strutbox
   \leftskip=\footindent \rightskip=\z@skip
   \parindent=0.5\footindent \parfillskip=0pt plus 1fil
   \spaceskip=\z@skip \xspaceskip=\z@skip
   \Textindent{$ #1 $}\footstrut\futurelet\next\fo@t}
\def\Textindent#1{\noindent\llap{#1\enspace}\ignorespaces}
\def\footnote#1{\attach{#1}\vfootnote{#1}}

\def\foot{\attach\footsymbolgen\vfootnote{\footsymbol}}
\let\footsymbol=\star
\newcount\lastf@@t           \lastf@@t=-1
\newcount\footsymbolcount    \footsymbolcount=0
\newif\ifPhysRev
\def\footsymbolgen{\relax \ifPhysRev \iffrontpage \NPsymbolgen\else
      \PRsymbolgen\fi \else \NPsymbolgen\fi
   \global\lastf@@t=\pageno \footsymbol }
\def\NPsymbolgen{\ifnum\footsymbolcount<0 \global\footsymbolcount=0\fi
   {\iffrontpage \else \advance\lastf@@t by 1 \fi
    \ifnum\lastf@@t<\pageno \global\footsymbolcount=0
     \else \global\advance\footsymbolcount by 1 \fi }
   \ifcase\footsymbolcount \fd@f\star\or \fd@f\dagger\or \fd@f\ast\or
    \fd@f\ddagger\or \fd@f\natural\or \fd@f\diamond\or \fd@f\bullet\or
    \fd@f\nabla\else \fd@f\dagger\global\footsymbolcount=0 \fi }
\def\fd@f#1{\xdef\footsymbol{#1}}
\def\PRsymbolgen{\ifnum\footsymbolcount>0 \global\footsymbolcount=0\fi
      \global\advance\footsymbolcount by -1
      \xdef\footsymbol{\sharp\number-\footsymbolcount} }
\def\space@ver#1{\let\@sf=\empty \ifmmode #1\else \ifhmode
   \edef\@sf{\spacefactor=\the\spacefactor}\unskip${}#1$\relax\fi\fi}
\def\attach#1{\space@ver{\strut^{\mkern 2mu #1} }\@sf\ }
\def\atttach#1{\space@ver{\strut{\mkern 2mu #1} }\@sf\ }
%
%
%
\newcount\chapternumber      \chapternumber=0
\newcount\sectionnumber      \sectionnumber=0
\newcount\equanumber         \equanumber=0
\let\chapterlabel=0
\newtoks\chapterstyle        \chapterstyle={\Number}
\newskip\chapterskip         \chapterskip=\bigskipamount
\newskip\sectionskip         \sectionskip=\medskipamount
\newskip\headskip            \headskip=8pt plus 3pt minus 3pt
\newdimen\chapterminspace    \chapterminspace=15pc
\newdimen\sectionminspace    \sectionminspace=10pc
\newdimen\referenceminspace  \referenceminspace=25pc
\def\chapterreset{\global\advance\chapternumber by 1
   \ifnum\the\equanumber<0 \else\global\equanumber=0\fi
   \sectionnumber=0 \makel@bel}
\def\makel@bel{\xdef\chapterlabel{%
\the\chapterstyle{\the\chapternumber}.}}
\def\sectionlabel{\number\sectionnumber \quad }
\def\alphabetic#1{\count255='140 \advance\count255 by #1\char\count255}
\def\Alphabetic#1{\count255='100 \advance\count255 by #1\char\count255}
\def\Roman#1{\uppercase\expandafter{\romannumeral #1}}
\def\roman#1{\romannumeral #1}
\def\Number#1{\number #1}
\def\unnumberedchapters{\let\makel@bel=\relax \let\chapterlabel=\relax
\let\sectionlabel=\relax \equanumber=-1 }
\def\titlestyle#1{\par\begingroup \interlinepenalty=9999
     \leftskip=0.02\hsize plus 0.23\hsize minus 0.02\hsize
     \rightskip=\leftskip \parfillskip=0pt
     \hyphenpenalty=9000 \exhyphenpenalty=9000
     \tolerance=9999 \pretolerance=9000
     \spaceskip=0.333em \xspaceskip=0.5em
     \iftwelv@\fourteenpoint\else\twelvepoint\fi
   \noindent #1\par\endgroup }
\def\spacecheck#1{\dimen@=\pagegoal\advance\dimen@ by -\pagetotal
   \ifdim\dimen@<#1 \ifdim\dimen@>0pt \vfil\break \fi\fi}
\def\chapter#1{\par \penalty-300 \vskip\chapterskip
   \spacecheck\chapterminspace
   \chapterreset \titlestyle{\chapterlabel \ #1}
   \nobreak\vskip\headskip \penalty 30000
   \wlog{\string\chapter\ \chapterlabel} }

\def\section#1{\par \ifnum\the\lastpenalty=30000\else
   \penalty-200\vskip\sectionskip \spacecheck\sectionminspace\fi
   \wlog{\string\section\ \chapterlabel \the\sectionnumber}
   \global\advance\sectionnumber by 1  \noindent
   {\caps\enspace\chapterlabel \sectionlabel #1}\par
   \nobreak\vskip\headskip \penalty 30000 }
\def\subsection#1{\par
   \ifnum\the\lastpenalty=30000\else \penalty-100\smallskip \fi
   \noindent\undertext{#1}\enspace \vadjust{\penalty5000}}

\def\undertext#1{\vtop{\hbox{#1}\kern 1pt \hrule}}
\def\ack{\par\penalty-100\medskip \spacecheck\sectionminspace
   \line{\fourteenrm\hfil ACKNOWLEDGEMENTS\hfil}\nobreak\vskip\headskip }
\def\APPENDIX#1#2{\par\penalty-300\vskip\chapterskip
   \spacecheck\chapterminspace \chapterreset \xdef\chapterlabel{#1}
   \titlestyle{APPENDIX #2} \nobreak\vskip\headskip \penalty 30000
   \wlog{\string\Appendix\ \chapterlabel} }
\def\Appendix#1{\APPENDIX{#1}{#1}}
\def\appendix{\APPENDIX{A}{}}
%
%
%
\def\eqname#1{\relax \ifnum\the\equanumber<0
     \xdef#1{{\rm(\number-\equanumber)}}\global\advance\equanumber by -1
    \else \global\advance\equanumber by 1
      \xdef#1{{\rm(\chapterlabel \number\equanumber)}} \fi}

\def\eqn#1{\eqno\eqname{#1}#1}

\def\eqinsert#1{\noalign{\dimen@=\prevdepth \nointerlineskip
   \setbox0=\hbox to\displaywidth{\hfil #1}
   \vbox to 0pt{\vss\hbox{$\!\box0\!$}\kern-0.5\baselineskip}
   \prevdepth=\dimen@}}
\def\sequentialequations{\globaleqnumbers}
%
%
\def\GENITEM#1;#2{\par \hangafter=0 \hangindent=#1
    \Textindent{$ #2 $}\ignorespaces}
\outer\def\newitem#1=#2;{\gdef#1{\GENITEM #2;}}
\newdimen\itemsize                \itemsize=30pt
\newitem\item=1\itemsize;
\newitem\sitem=1.75\itemsize;     
\newitem\ssitem=2.5\itemsize;     
\outer\def\newlist#1=#2&#3&#4;{\toks0={#2}\toks1={#3}%
   \count255=\escapechar \escapechar=-1
   \alloc@0\list\countdef\insc@unt\listcount     \listcount=0
   \edef#1{\par
      \countdef\listcount=\the\allocationnumber
      \advance\listcount by 1
      \hangafter=0 \hangindent=#4
      \Textindent{\the\toks0{\listcount}\the\toks1}}
   \expandafter\expandafter\expandafter
    \edef\c@t#1{begin}{\par
      \countdef\listcount=\the\allocationnumber \listcount=1
      \hangafter=0 \hangindent=#4
      \Textindent{\the\toks0{\listcount}\the\toks1}}
   \expandafter\expandafter\expandafter
    \edef\c@t#1{con}{\par \hangafter=0 \hangindent=#4 \noindent}
   \escapechar=\count255}
\def\c@t#1#2{\csname\string#1#2\endcsname}
\newlist\point=\Number&.&1.0\itemsize;
\newlist\subpoint=(\alphabetic&)&1.75\itemsize;
\newlist\subsubpoint=(\roman&)&2.5\itemsize;
\newlist\cpoint=\Roman&.&1.0\itemsize;
%

%
%
%
\newcount\referencecount     \referencecount=0
\newif\ifreferenceopen       \newwrite\referencewrite
\newtoks\rw@toks
\def\NPrefmark#1{\atttach{\rm [ #1 ] }}
\let\PRrefmark=\attach
\def\CErefmark#1{\attach{\scriptstyle  #1 ) }}
\def\refmark#1{\relax\ifPhysRev\PRrefmark{#1}\else\NPrefmark{#1}\fi}
\def\crefmark#1{\relax\CErefmark{#1}}
\def\refend{\refmark{\number\referencecount}}
\newcount\lastrefsbegincount \lastrefsbegincount=0
\def\refsend{\refmark{\count255=\referencecount
   \advance\count255 by-\lastrefsbegincount
   \ifcase\count255 \number\referencecount
   \or \number\lastrefsbegincount,\number\referencecount
   \else \number\lastrefsbegincount-\number\referencecount \fi}}
\def\crefsend{\crefmark{\count255=\referencecount
   \advance\count255 by-\lastrefsbegincount
   \ifcase\count255 \number\referencecount
   \or \number\lastrefsbegincount,\number\referencecount
   \else \number\lastrefsbegincount-\number\referencecount \fi}}
\def\refch@ck{\chardef\rw@write=\referencewrite
   \ifreferenceopen \else \referenceopentrue
   \immediate\openout\referencewrite=referenc.texauxil \fi}
%
{\catcode`\^^M=\active 
  \gdef\obeyendofline{\catcode`\^^M\active \let^^M\ }}%
%
{\catcode`\^^M=\active 
  \gdef\ignoreendofline{\catcode`\^^M=5}}
{\obeyendofline\gdef\rw@start#1{\def\t@st{#1} \ifx\t@st\blankend%
\endgroup \@sf \relax \else \ifx\t@st\bl@nkend \endgroup \@sf \relax%
\else \rw@begin#1
\backtotext
\fi \fi } }
{\obeyendofline\gdef\rw@begin#1
{\def\n@xt{#1}\rw@toks={#1}\relax%
\rw@next}}
\def\blankend{}
{\obeylines\gdef\bl@nkend{
}}
\newif\iffirstrefline  \firstreflinetrue
\def\rwr@teswitch{\ifx\n@xt\blankend \let\n@xt=\rw@begin %
 \else\iffirstrefline \global\firstreflinefalse%
\immediate\write\rw@write{\noexpand\obeyendofline \the\rw@toks}%
\let\n@xt=\rw@begin%
      \else\ifx\n@xt\rw@@d \def\n@xt{\immediate\write\rw@write{%
        \noexpand\ignoreendofline}\endgroup \@sf}%
             \else \immediate\write\rw@write{\the\rw@toks}%
             \let\n@xt=\rw@begin\fi\fi \fi}
\def\rw@next{\rwr@teswitch\n@xt}
\def\rw@@d{\backtotext} \let\rw@end=\relax
\let\backtotext=\relax

\newdimen\refindent     \refindent=30pt
\def\refitem#1{\par \hangafter=0 \hangindent=\refindent \Textindent{#1}}
\def\REFNUM#1{\space@ver{}\refch@ck \firstreflinetrue%
 \global\advance\referencecount by 1 \xdef#1{\the\referencecount}}
\def\refnum#1{\space@ver{}\refch@ck \firstreflinetrue%
 \global\advance\referencecount by 1 \xdef#1{\the\referencecount}\refend}

\def\REF#1{\REFNUM#1%
 \immediate\write\referencewrite{%
 \noexpand\refitem{#1.}}%
\begingroup\obeyendofline\rw@start}
\def\ref{\refnum\?%
 \immediate\write\referencewrite{\noexpand\refitem{\?.}}%
\begingroup\obeyendofline\rw@start}
\def\Ref#1{\refnum#1%
 \immediate\write\referencewrite{\noexpand\refitem{#1.}}%
\begingroup\obeyendofline\rw@start}
\def\REFS#1{\REFNUM#1\global\lastrefsbegincount=\referencecount
\immediate\write\referencewrite{\noexpand\refitem{#1.}}%
\begingroup\obeyendofline\rw@start}
\newcount\figurecount     \figurecount=0
\newif\iffigureopen       \newwrite\figurewrite
\def\figch@ck{\chardef\rw@write=\figurewrite \iffigureopen\else
   \immediate\openout\figurewrite=figures.texauxil
   \figureopentrue\fi}
\def\FIGNUM#1{\space@ver{}\figch@ck \firstreflinetrue%
 \global\advance\figurecount by 1 \xdef#1{\the\figurecount}}
\def\FIG#1{\FIGNUM#1
   \immediate\write\figurewrite{\noexpand\refitem{#1.}}%
   \begingroup\obeyendofline\rw@start}
\def\fig{\FIGNUM\? fig.~\?%
\immediate\write\figurewrite{\noexpand\refitem{\?.}}%
\begingroup\obeyendofline\rw@start}
\def\figure{\FIGNUM\? figure~\?
   \immediate\write\figurewrite{\noexpand\refitem{\?.}}%
   \begingroup\obeyendofline\rw@start}
\def\Fig{\FIGNUM\? Fig.~\?%
\immediate\write\figurewrite{\noexpand\refitem{\?.}}%
\begingroup\obeyendofline\rw@start}
\def\Figure{\FIGNUM\? Figure~\?%
\immediate\write\figurewrite{\noexpand\refitem{\?.}}%
\begingroup\obeyendofline\rw@start}
\newcount\tablecount     \tablecount=0
\newif\iftableopen       \newwrite\tablewrite
\def\tabch@ck{\chardef\rw@write=\tablewrite \iftableopen\else
   \immediate\openout\tablewrite=tables.texauxil
   \tableopentrue\fi}
\def\TABNUM#1{\space@ver{}\tabch@ck \firstreflinetrue%
 \global\advance\tablecount by 1 \xdef#1{\the\tablecount}}
\def\TABLE#1{\TABNUM#1
   \immediate\write\tablewrite{\noexpand\refitem{#1.}}%
   \begingroup\obeyendofline\rw@start}
\def\Table{\TABNUM\? Table~\?%
\immediate\write\tablewrite{\noexpand\refitem{\?.}}%
\begingroup\obeyendofline\rw@start}
%

%
%
%
\def\masterreset{\global\pagenumber=1 \global\chapternumber=0
   \ifnum\the\equanumber<0\else \global\equanumber=0\fi
   \global\sectionnumber=0
   \global\referencecount=0 \global\figurecount=0 \global\tablecount=0 }
\def\FRONTPAGE{\ifvoid255\else\vfill\penalty-2000\fi
      \masterreset\global\frontpagetrue
      \global\lastf@@t=0 \global\footsymbolcount=0}

\def\paperstyle{\letterstylefalse\normalspace\papersize}
\def\letterstyle{\letterstyletrue\singlespace\lettersize}
\def\papersize{\hsize=6.5truein\vsize=9.1truein\hoffset=-.3truein
               \voffset=-.4truein\skip\footins=\bigskipamount}
\def\lettersize{\hsize=6.5truein\vsize=9.1truein\hoffset=-.3truein
    \voffset=.1truein\skip\footins=\smallskipamount \multiply
    \skip\footins by 3 }
\paperstyle   
%
%
\def\MEMO{\letterstyle\FRONTPAGE \letterfrontheadline={\hfil}
    \line{\quad\fourteenrm CERN MEMORANDUM\hfil\twelverm\the\date\quad}
    \medskip \memod@f}

\def\memit@m#1{\smallskip \hangafter=0 \hangindent=1in
      \Textindent{\caps #1}}
\def\memod@f{\xdef\mto{\memit@m{To:}}\xdef\from{\memit@m{From:}}%
     \xdef\topic{\memit@m{Topic:}}\xdef\subject{\memit@m{Subject:}}%
     \xdef\rule{\bigskip\hrule height 1pt\bigskip}}
\memod@f
\newskip\lettertopfil
\lettertopfil = 0pt plus 1.5in minus 0pt
\newskip\letterbottomfil
\letterbottomfil = 0pt plus 2.3in minus 0pt
\newskip\spskip \setbox0\hbox{\ } \spskip=-1\wd0
\def\addressee#1{\medskip\rightline{\the\date\hskip 30pt} \bigskip
   \vskip\lettertopfil
   \ialign to\hsize{\strut ##\hfil\tabskip 0pt plus \hsize \cr #1\crcr}
   \medskip\noindent\hskip\spskip}
\newskip\signatureskip       \signatureskip=40pt
\def\signed#1{\par \penalty 9000 \bigskip \dt@pfalse
  \everycr={\noalign{\ifdt@p\vskip\signatureskip\global\dt@pfalse\fi}}
  \setbox0=\vbox{\singlespace \halign{\tabskip 0pt \strut ##\hfil\cr
   \noalign{\global\dt@ptrue}#1\crcr}}
  \line{\hskip 0.5\hsize minus 0.5\hsize \box0\hfil} \medskip }

\def\endletter{\ifnum\pagenumber=1 \vskip\letterbottomfil\supereject
\else \vfil\supereject \fi}
\newbox\letterb@x
\def\lettertext{\par\unvcopy\letterb@x\par}
\def\multiletter{\setbox\letterb@x=\vbox\bgroup
      \everypar{\vrule height 1\baselineskip depth 0pt width 0pt }
      \singlespace \topskip=\baselineskip }
\def\letterend{\par\egroup}
%
%
%
\newskip\frontpageskip
\newtoks\pubtype
\newtoks\Pubnum
\newtoks\pubnum
\newtoks\pubnu
\newtoks\pubn
\newif\ifp@bblock  \p@bblocktrue
\def\PH@SR@V{\doubl@true \baselineskip=24.1pt plus 0.2pt minus 0.1pt
             \parskip= 3pt plus 2pt minus 1pt }
\def\PHYSREV{\paperstyle\PhysRevtrue\PH@SR@V}
\def\titlepage{\FRONTPAGE\paperstyle\ifPhysRev\PH@SR@V\fi
   \ifp@bblock\p@bblock\fi}
\def\nopubblock{\p@bblockfalse}
\def\endpage{\vfil\break}
\frontpageskip=1\medskipamount plus .5fil
\pubtype={\tensl Preliminary Version}
\Pubnum={$\rm CERN-TH.\the\pubnum $}
\pubnum={0000}
\def\p@bblock{\begingroup \tabskip=\hsize minus \hsize
   \baselineskip=1.5\ht\strutbox \topspace-2\baselineskip
   \halign to\hsize{\strut ##\hfil\tabskip=0pt\crcr
   \the \pubn\cr
   \the \Pubnum\cr
   \the \pubnu\cr
   \the \date\cr}\endgroup}
\def\title#1{\vskip\frontpageskip \titlestyle{#1} \vskip\headskip }
\def\author#1{\vskip\frontpageskip\titlestyle{\twelvecp #1}\nobreak}

\def\address#1{\par\kern 5pt\titlestyle{\twelvepoint\it #1}}
\def\andaddress{\par\kern 5pt \centerline{\sl and} \address}

\def\abstract{\vskip\frontpageskip\centerline{\fourteenrm ABSTRACT}
              \vskip\headskip }

%
%
%

\def\\{\relax\ifmmode\backslash\else$\backslash$\fi}
\def\globaleqnumbers{\relax\ifnum\the\equanumber<0%
\else\global\equanumber=-1\fi}

\def\journal#1&#2(#3){\unskip, \sl #1~\bf #2 \rm (19#3) }

\def\topspace{\hrule height 0pt depth 0pt \vskip}

\let\int=\intop         
\def\prop{\mathrel{{\mathchoice{\pr@p\scriptstyle}{\pr@p\scriptstyle}{
                \pr@p\scriptscriptstyle}{\pr@p\scriptscriptstyle} }}}
\def\pr@p#1{\setbox0=\hbox{$\cal #1 \char'103$}
   \hbox{$\cal #1 \char'117$\kern-.4\wd0\box0}}
\def\lsim{\mathrel{\mathpalette\@versim<}}
\def\gsim{\mathrel{\mathpalette\@versim>}}
\def\@versim#1#2{\lower0.2ex\vbox{\baselineskip\z@skip\lineskip\z@skip
  \lineskiplimit\z@\ialign{$\m@th#1\hfil##\hfil$\crcr#2\crcr\sim\crcr}}}
\def\leftrightarrowfill{$\m@th \mathord- \mkern-6mu
        \cleaders\hbox{$\mkern-2mu \mathord- \mkern-2mu$}\hfil
        \mkern-6mu \mathord\leftrightarrow$}
\def\lrover#1{\vbox{\ialign{##\crcr
        \leftrightarrowfill\crcr\noalign{\kern-1pt\nointerlineskip}
        $\hfil\displaystyle{#1}\hfil$\crcr}}}
%
%
%
\let\sec@nt=\sec
\def\sec{\relax\ifmmode\let\n@xt=\sec@nt\else\let\n@xt\section\fi\n@xt}
\def\obsolete#1{\message{Macro \string #1 is obsolete.}}
\def\firstsec#1{\obsolete\firstsec \section{#1}}
\def\firstsubsec#1{\obsolete\firstsubsec \subsection{#1}}
\def\thispage#1{\obsolete\thispage \global\pagenumber=#1\frontpagefalse}
\def\thischapter#1{\obsolete\thischapter \global\chapternumber=#1}
\def\nextequation#1{\obsolete\nextequation \global\equanumber=#1
   \ifnum\the\equanumber>0 \global\advance\equanumber by 1 \fi}
\def\BOXITEM{\afterassigment\B@XITEM\setbox0=}
\def\B@XITEM{\par\hangindent\wd0 \noindent\box0 }
%

%
%

%
%

%
%

%

%

%

%

%

%
%
%
\def\boxit#1{\vbox{\hrule\hbox{\vrule\kern3pt\vbox{\kern3pt#1\kern3pt}
\kern3pt\vrule}\hrule}}
%
%
%
\catcode`@=12 
\message{ by V.K./U.B.}
\everyjob{\input imyphyx }
%
%
%
%
%
%
%
%
%
%
%
%
%
\catcode`@=11

\font\seventeencp=cmcsc10 scaled\magstep3
\def\SIZE{\hsize=6.6truein\vsize=9.1truein}
\def\OFFSET{\voffset=1.2truein\hoffset=.8truein}
\def\papersize{\SIZE\OFFSET\skip\footins=\bigskipamount
\normaldisplayskip= 30pt plus 5pt minus 10pt}
\Pubnum={\rm CERN$-$TH.\the\pubnum }
\def\title#1{\vskip\frontpageskip\vskip .50truein
     \titlestyle{\seventeencp #1} \vskip\headskip\vskip\frontpageskip
     \vskip .2truein}
\def\author#1{\vskip .27truein\titlestyle{#1}\nobreak}

\def\p@bblock{\begingroup \tabskip=\hsize minus \hsize
   \baselineskip=1.5\ht\strutbox \topspace-2\baselineskip
   \halign to\hsize{\strut ##\hfil\tabskip=0pt\crcr
   \the \Pubnum\cr}\endgroup}
\def\makefootline{\iffrontpage\vskip .27truein\line{\the\footline}
                 \vskip -.1truein\line{\the\date\hfil}
              \else\line{\the\footline}\fi}
\paperfootline={\iffrontpage \the\Pubnum\hfil\else\hfil\fi}
\paperheadline={\iffrontpage\hfil
                \else\twelverm\hss $-$\ \folio\ $-$\hss\fi}
\newif\ifmref  
\newif\iffref  
\def\xrefsend{\xrefmark{\count255=\referencecount
\advance\count255 by-\lastrefsbegincount
\ifcase\count255 \number\referencecount
\or \number\lastrefsbegincount,\number\referencecount
\else \number\lastrefsbegincount-\number\referencecount \fi}}
\def\xrefsdub{\xrefmark{\count255=\referencecount
\advance\count255 by-\lastrefsbegincount
\ifcase\count255 \number\referencecount
\or \number\lastrefsbegincount,\number\referencecount
\else \number\lastrefsbegincount,\number\referencecount \fi}}
\def\xREFNUM#1{\space@ver{}\refch@ck\firstreflinetrue%
\global\advance\referencecount by 1
\xdef#1{\xrefend}}
\def\xrefend{\xrefmark{\number\referencecount}}
\def\xrefmark#1{[{#1}]}
\def\xRef#1{\xREFNUM#1\immediate\write\referencewrite%
{\noexpand\refitem{#1}}\begingroup\obeyendofline\rw@start}%
\def\xREFS#1{\xREFNUM#1\global\lastrefsbegincount=\referencecount%
\immediate\write\referencewrite{\noexpand\refitem{#1}}%
\begingroup\obeyendofline\rw@start}
\def\rrr#1#2{\relax\ifmref{\iffref\xREFS#1{#2}%
\else\xRef#1{#2}\fi}\else\xRef#1{#2}\xrefend\fi}
\def\multref#1#2{\mreftrue\freftrue{#1}%
\freffalse{#2}\mreffalse\xrefsend}
\referencecount=0
%
\space@ver{}\refch@ck\firstreflinetrue%
\immediate\write\referencewrite{}%
\begingroup\obeyendofline\rw@start{}%
\def\plb#1({Phys.\ Lett.\ $\underline  {#1B}$\ (}
\def\nup#1({Nucl.\ Phys.\ $\underline {B#1}$\ (}
\def\plt#1({Phys.\ Lett.\ $\underline  {B#1}$\ (}
\def\cmp#1({Comm.\ Math.\ Phys.\ $\underline  {#1}$\ (}
\def\prp#1({Phys.\ Rep.\ $\underline  {#1}$\ (}
\def\prl#1({Phys.\ Rev.\ Lett.\ $\underline  {#1}$\ (}
\def\prv#1({Phys.\ Rev. $\underline  {D#1}$\ (}
\def\und#1({            $\underline  {#1}$\ (}
\message{ by W.L.}
\everyjob{\input offset }
\catcode`@=12

\let\it=\sl

%

%
\def\OFFSET{\hoffset=6.pt\voffset=40.pt}
\def\SIZE{\hsize=420.pt\vsize=620.pt}
\OFFSET

\def\ABK{\rrr\ABK{I. Antoniadis, C. Bachas and C. Kounnas,
        \nup289 (1987) 87.}}

\def\KAPLU{\rrr\KAPLU{V. Kaplunovsky, \nup307 (1988) 145.}}

\def\REVDU{\rrr\REVDU{For a review see: D. L\"ust, preprints
CERN-TH.6143/91, 6259/91; L. Iba\~nez, preprint CERN-TH.6342/91.}}

\def\TWOGAU{\rrr\TWOGAU{N.V. Krasnikov, \plt193 (1987) 37;
L. Dixon, talk presented at the A.P.S. D.P.F. Meeting at Houston (1990);
V. Kaplunovsky, talk presented at the `Strings 90' workshop at
College Station (1990); T.R. Taylor, \plt252 (1990) 59.}}

\def\SDUAL{\rrr\SDUAL{A. Font, L.E. Ib\'a\~nez, D. L\"ust and F. Quevedo,
\plt249 (1990) 35.}}

\def\HIGGS{\rrr\HIGGS{I. Antoniadis, C. Bachas, C. Kounnas, \plt200
(1988) 297;
L.E. Ib\'a\~nez, W. Lerche, D. L\"ust and
S. Theisen, \nup352 (1991) 435.}}

\def\NIL{\rrr\NIL{H.-P. Nilles, \plb180 (1986) 240;
M.A. Shifman and A.I. Vainshtein, \nup359
   (1991) 571;
J.A. Casas and C. Mu\~noz, \plt271 (1991) 85.}}

\def\GRA{\rrr\GRA{G.G. Ross, \plt211 (1988) 315.}}

\def\KLN{\rrr\KLN{I. Antoniadis, J. Ellis, A.B. Lahanas and
D.V. Nanopoulos, \plt241 (1990) 24;
S. Kalara, J.L. Lopez and D.V. Nanopoulos,
{\it ``Gauge and Matter Condensates in realistic String Models'',}
  preprint CTP-TAMU-69/91 (1991).}}

\def\HAVA{\rrr\HAVA{S. Hamidi and C. Vafa, \nup279 (1987) 465; L. Dixon,
D. Friedan, E. Martinec and S. Shenker, \nup 282 (1987) 13.}}

\def\IBLUA{\rrr\IBLUA{L. Iba\~nez and D. L\"ust, CERN preprint
to appear;
V. Kaplunovsky and J. Louis, preprint to appear.}}

\def\LT{\rrr\LT{D. L\"ust and T.R. Taylor, \plt253 (1991) 335.}}

\def\CGM{\rrr\CGM{J.A. Casas, F. Gomez and C. Mu\~noz, {\it ``Complete
   Structure of ${\bf Z}_n$ Yukawa Couplings'',} preprint
CERN-TH.6194/91 (1991).}}

\def\CM{\rrr\CM{B. de Carlos,
J.A. Casas and C. Mu\~noz, \plt263 (1991) 248.}}

\def\CMA{\rrr\CMA{B. de Carlos,
J.A. Casas and C. Mu\~noz, CERN preprint to appear.}}

\def\CLMR{\rrr\CLMR{J.A. Casas, Z. Lalak, C. Mu\~noz and G.G. Ross,
\nup347 (1990) 243.}}

\def\CREMMER{\rrr\CREMMER{E. Cremmer, S. Ferrara, L. Girardello and
          A. Van Proeyen, \nup212 (1983) 413.}}

\def\DFKZ{\rrr\DFKZ{J.P. Derendinger, S. Ferrara, C. Kounnas and F. Zwirner,
         {\it ``On loop corrections to string effective field theories:
         field-dependent gauge couplings and sigma-model anomalies'',}
        preprint CERN-TH.6004/91, LPTENS 91-4 (revised version) (1991);
\plb271 (1991) 307                .}}

\def\KATSUKI{\rrr\KATSUKI{Y. Katsuki, Y. Kawamura, T. Kobayashi, Y. Ono
and K. Tanioka, \plt218 (1989) 169.}}

\def\ORBIMA{\rrr\ORBIMA{J.A. Casas, A. de la Macorra, M. Mondragon
and C. Mu\~noz, \plt247 (1990) 50.}}

\def\ORBIM{\rrr\ORBIM{J.A. Casas, E.K. Katehou and C. Mu\~noz, \nup317
(1989) 171; J.A. Casas and C. Mu\~noz, \plt214 (1988) 63;
A. Font, L. Iba\~nez, H.-P. Nilles and F. Quevedo, \plt210 (1988) 101.}}

\def\ORBICON{\rrr\ORBICON{S. Mahapatra, \plt223 (1989) 47;
Y. Katsuki, Y. Kawamura, T. Kobayashi, N. Ohtsubo, Y. Ono and K. Tanioka,
\nup341 (1990) 611.}}

\def\IBPR{\rrr\IBPR{L. Ib\'a\~nez, unpublished notes.}}

\def\FLIP{\rrr\FLIP{I. Antoniadis, J. Ellis, J.S. Hagelin and
D.V. Nanopoulos, \plt205 (1988) 459, \plt213 (1988) 56.}}

\def\ANTO{\rrr\ANTO{I. Antoniadis, \plt246 (1990) 377.}}

\def\LOUIS{\rrr\LOUIS{J. Louis, {\it
         ``Non-harmonic gauge coupling constants in supersymmetry
         and superstring theory'',} preprint SLAC-PUB-5527 (1991).}}

\def\ORBI{\rrr\ORBI{L. Dixon, J. Harvey, C.~Vafa and E.~Witten,
         \nup261 (1985) 651, \nup274 (1986) 285;
                    L.E. Ib\'a\~nez, H.P. Nilles and F. Quevedo,
                     \plt187 (1987) 25.}}

\def\DKLB{\rrr\DKLB{L. Dixon, V. Kaplunovsky and J. Louis,
         \nup355 (1991) 649.}}

\def\DKLA{\rrr\DKLA{L. Dixon, V. Kaplunovsky and J. Louis, \nup329 (1990)
            27.}}

\def\FKLZ{\rrr\FKLZ{S. Ferrara, C. Kounnas, D. L\"ust and F. Zwirner,
           {\it ``Duality Invariant Partition Functions and
           Automorphic Superpotentials for (2,2) String
           Compactifications''}, preprint CERN-TH.6090/91 (1991),
           to appear in Nucl. Phys. B.}}

\def\FILQ{\rrr\FILQ{A. Font, L.E. Ib\'a\~nez, D. L\"ust and F. Quevedo,
           \plt245 (1990) 401.}}

\def\FMTV{\rrr\FMTV{S. Ferrara, N. Magnoli, T.R. Taylor and
           G. Veneziano, \plt245 (1990) 409.}}

\def\FILQ{\rrr\FILQ{A. Font, L.E. Ib\'a\~nez, D. L\"ust and F. Quevedo,
           \plt245 (1990) 401.}}

\def\OTHER{\rrr\OTHER{
           H.P. Nilles and M.
           Olechowski, \plt248 (1990) 268; P. Binetruy and M.K.
           Gaillard, \plt253 (1991) 119; M. Cvetic, A. Font, L.E.
           Ib\'a\~nez, D. L\"ust and F. Quevedo, \nup361 (1991) 194;
           J. Louis, {\it ``Status of Supersymmetry Breaking
           in String Theory'',} SLAC-PUB-5645 (1991).}}

\def\FIQ{\rrr\FIQ{A. Font, L.E. Ib\'a\~nez and F. Quevedo,
        \plt217 (1989) 272.}}

\def\FLST{\rrr\FLST{S. Ferrara,
         D. L\"ust, A. Shapere and S. Theisen, \plt225 (1989) 363.}}

\def\FLT{\rrr\FLT
{S. Ferrara, D. L\"ust and S. Theisen, \plt233 (1989) 147.}}

\def\CAOV{\rrr\CAOV{G. Lopes Cardoso and B. Ovrut, {\it ``A Green-Schwarz
            Mechanism for D=4, N=1 supergravity Anomalies,''}
            preprint UPR-0464T (1991); {\it ``Sigma Model Anomalies,
           Non-Harmonic gauge Couplings and String Theory,''}
           preprint UPR-0481T (1991).
            }}

\def\IBLU{\rrr\IBLU{L. Ib\'a\~nez and D. L\"ust, \plt267 (1991) 51.}}

\def\ANT{\rrr\ANT{I. Antoniadis, K.S. Narain and T.R. Taylor,
        \plt267 (1991) 37.}}

\def\GS{\rrr\GS{M.B. Green and J.H. Schwarz, \plb149 (1984) 117.}}

\def\WI{\rrr\WI{E. Witten, \plb155 (1985) 151; S. Ferrara, C. Kounnas
and M. Porrati, \plt181 (1986) 263; M. Cvetic, J. Louis and B. Ovrut,
\plt206 (1988) 227.}}

\def\OFFSET{\hoffset=12.pt\voffset=55.pt}
\def\SIZE{\hsize=420.pt\vsize=620.pt}

\catcode`@=12
\newtoks\Pubnumtwo
\newtoks\Pubnumthree
\catcode`@=11
\def\p@bblock{\begingroup\tabskip=\hsize minus\hsize
   \baselineskip=0.5\ht\strutbox\topspace-2\baselineskip
   \halign to \hsize{\strut ##\hfil\tabskip=0pt\crcr
   \the\Pubnum\cr  \the\Pubnumtwo\cr 
   \the\pubtype\cr}\endgroup}
\pubnum={6358/91}
\date{December 1991}
\pubtype={}
\titlepage
\vskip -.6truein
\title{Duality-Invariant Gaugino Condensation and
One-loop Corrected K\"ahler Potentials
in String Theory}
 \centerline{{\bf Dieter L\"ust}\foot{Heisenberg Fellow}}
 \vskip .1truein
\centerline{and}
   \vskip 0.1truein
  \centerline{\bf Carlos Mu\~noz}
  \vskip 0.1truein
  \centerline{CERN, 1211 Geneva 23, Switzerland}
\abstract\noindent\nobreak
The duality-invariant gaugino condensation with or without massive
matter fields is re-analysed, taking into account the
dependence of the string threshold corrections on the moduli fields
and recent results concerning one-loop corrected K\"ahler potentials.
The scalar potential of the theory for a generic superpotential
is also calculated.
\vskip 1.0cm
\endpage
\pagenumber=1
\sequentialequations

The understanding of supersymmetry breaking is crucial to make
contact between four-dimensional superstrings and low-energy physics.
Recently, the duality-invariant gaugino condensation
mechanism \multref\FILQ{\FMTV\OTHER}\ provided
substantial information about non-perturbative supersymmetry breaking
within the context of low-energy effective string Lagrangians.
In particular, important insight in
the related problem of the dynamical determination of the
moduli parameters, i.e.
the lifting of the huge vacuum degeneracy of string compactifications,
was gained.
These results are essentially based on the field-theory one-loop
running of the effective gauge coupling constant in the hidden sector,
focusing in addition on the
moduli-dependent string threshold effects \KAPLU,
which are due to the presence of heavy string modes. This analysis,
although leading to qualitatively correct results in most
of the known cases,
appears to be slightly incomplete, since
loop corrections to the K\"ahler
potential \DFKZ\ were not taken into account.
Specifically we will discuss the duality-invariant gaugino
condensation mechanism in a pure Yang--Mills hidden sector for
orbifold compactifications \ORBI. We will also calculate
the scalar potential of the theory
for a generic superpotential using the one-loop K\"ahler function.
As a result we will find that the
use of the loop corrected K\"ahler potential can be regarded
as a simple redefinition of the tree-level gauge coupling constant.
In turn this redefinition will not qualitatively
affect the previous results about the dynamical supersymmetry
breaking and the determination of the moduli parameters, except
when those moduli are associated with complex planes rotated by all
orbifold twists. In these cases, to achieve the final goal that
all moduli acquire expectation values dynamically, determining
the size of the compactified space, one should also include \GRA,\CLMR\
the moduli-dependent (twisted) Yukawa
couplings as an extra piece in the superpotential.
Finally, we will discuss  duality-invariant gaugino condensation
in hidden sectors with massive matter fields. This was previously
studied in refs.\LT\ and \CM\
only for untwisted matter fields and
without taking into account the above-mentioned
corrections. As for the pure gauge case, the qualitative results are
not affected, except for the fact that for twisted massive matter fields
the moduli-dependent Yukawa couplings explicitly appear
in the gaugino condensate.

Let us start by recalling some recent results on the effective $N=1$
supergravity Lagrangian for orbifold compactifications.
Apart from the gravitational supermultiplet
and gauge vector supermultiplets
of the gauge group $G$,
we first consider as the relevant massless string degrees of freedom
the dilaton--axion chiral field $S$ and the three internal
moduli fields $T_i$ ($i=1,2,3$), whose real parts
determine the sizes of the three underlying two-dimensional complex
planes and
whose imaginary parts are given by three internal axion fields:
$T_i=R^2_i+iB_i$. These three moduli fields are present in any
Abelian ${\bf Z}_M$ and ${\bf Z}_M\times {\bf Z}_N$ orbifold
compactification.
Finally we consider also additional matter fields $C_\alpha$
in the $\underline R_\alpha$ representation of $G$, which can
be either massless or massive (depending on the form of the
superpotential to be discussed in the following).

The couplings of the chiral fields
are determined by the real function $G(\phi,\bar\phi)$ \CREMMER, which is
a combination of the K\"ahler potential and the holomorphic
superpotential: $G(\phi,\bar \phi)=K(\phi,\bar \phi)+\log |W(\phi )|^2$.
At string tree level, the K\"ahler potential for $S,T$ and $C_\alpha$
has the simple form (at lowest order in the matter fields) \WI,\DKLA:
$K_{\rm tree}=-\log(S+\bar S)-\sum_{i=1}^3\log(T_i+\bar T_i)+
\sum_\alpha C_\alpha\bar C_\alpha \prod_{i=1}^3
(T_i+\bar T_i)^{n_\alpha^i}$.
The lowest order superpotential is cubic in the matter
fields:
$W_{\rm tree}=h_{\alpha\beta\gamma}(T_i)C_\alpha C_\beta C_\gamma$,
where the moduli-dependent functions $h_{\alpha\beta\gamma}(T_i)$
\HAVA\
are called Yukawa
couplings. Specifically, the Yukawa couplings $h_{\alpha\beta\gamma}$
are moduli-dependent functions if all fields $C$ are in the twisted
sectors of the orbifold, otherwise $h_{\alpha\beta\gamma}={\rm const}$.
For all matter fields  being untwisted,
the Yukawa couplings are non-zero if the fields belong to
different complex planes.
Finally, the couplings of the chiral fields to gauge
vector fields is determined
by the gauge kinetic function $f(\phi)$. At string tree
level this function is just given by the $S$-field, $f=S$, such
that the tree level gauge coupling constant of $G$ is determined
by the vacuum expectation value of the
dilaton field (we assume the Kac--Moody level to be one):
${1\over g_{\rm tree}^2}={S+\bar S\over2}$.

As discussed in ref.
\REVDU\ the underlying (super) conformal field theories
are invariant under the target space modular transformations
acting on the complex moduli fields $T_i$ as
$T_i\rightarrow{a_iT_i-ib_i\over ic_iT_i+d_i}$,
with $a_i,b_i,c_i,d_i\in
{\bf Z}$, $a_id_i-b_ic_i=1$. These transformations include the
well-known duality transformations $R_i\rightarrow{1\over R_i}$ as
well as discrete shifts of the axion fields $B_i$. Thus, up to
permutation symmetries, the generalized target duality symmetries
$\Gamma$ are described by the product of three modular groups:
$\Gamma=\lbrack SL(2,{\bf Z})\rbrack^3$.
Also the matter fields transform in general non-trivially under
target space modular transformations like (up to a possible constant
matrix) \FLT:
$$C_{\alpha}\rightarrow
C_\alpha \prod_{i=1}^3(ic_iT_i+d_i)^{n_\alpha^i}.\eqn\mattrans
$$
Thus the numbers $n_\alpha^i$ are called the modular weights of the
matter fields. Specifically, untwisted matter fields associated to
the $j^{\rm th}$ complex plane, $C^{\rm untw}=C^j$,
have modular
weights $n^i_\alpha=-\delta^i_j$.
A detailed discussion about the range of values
of the $n^i$ corresponding to various types of twisted matter
fields can be found in refs. \DKLA,\IBLUA.

Since the spectrum and all interactions of the underlying
conformal field theories are target space modular-invariant
at each order in string perturbation theory, the
effective string action has to be modular-invariant as well.
As discussed in refs. \FLST\ and
\FLT\ the requirement of target space modular
invariance puts strong constraints on the form of the
low-energy supergravity action involving the moduli fields $T_i$,
providing a link to the theory of automorphic functions.
Since at tree level the S-field is invariant under target space modular
transformations, the change of the K\"ahler potential
under these transformations has the form
$$K_{\rm tree}\rightarrow K_{\rm tree}+\sum_{i=1}^3
\log |ic_iT_i+d_i|^2.\eqn\kptra
$$
In order to obtain invariant matter couplings, the superpotential
then has to transform as (up to a field independent phase) \FLST
$$W_{\rm tree}\rightarrow {W_{\rm tree}\over \prod_{i=1}^3
(ic_iT_i+d_i)}.\eqn\suppottra
$$
Therefore the Yukawa couplings have to transform as (up to possible
constant matrices)
$$h_{\alpha\beta\gamma}(T_i)\rightarrow h_{\alpha\beta\gamma}(T_i)
\prod_{i=1}^3(ic_iT_i+d_i)^{(-1-n_\alpha^i-n_\beta^i-n_\gamma^i)}.
\eqn\yuktrans
$$

As mentioned in the introduction, loop corrections will modify
in general the K\"ahler potential and the gauge coupling constant.
However at finite order in perturbation theory the superpotential
is expected to be unchanged. If this non-renormalization theorem holds,
it can be concluded that the modular transformation rule eq.\suppottra\
holds in every order of perturbation theory. Consequently, also the
K\"ahler potential has to transform in any order of perturbation theory
as is displayed in eq.\kptra. Furthermore, these transformation rules
will still hold when
taking into account the non-perturbative modification
of the superpotential due to the gaugino condensate.
Now let us discuss the one-loop modification of the K\"ahler
potential and the gauge coupling constant for the case of a
pure Yang--Mills gauge theory.
Specifically, the one-loop K\"ahler potential is given by \DFKZ
$$\eqalign{K_{\rm 1-loop}&
=-\log\biggl\lbrack(S+\bar S)-{1\over 8\pi^2}\sum_{i=1}^3
\delta_{GS}^i\log(T_i+\bar T_i)\biggr\rbrack
-\sum_{i=1}^3\log(T_i+\bar T_i)\cr
&=-\log Y-\sum_{i=1}^3\log(T_i+\bar T_i).\cr}
\eqn\kploop
$$
$K_{\rm 1-loop}$ now leads to a mixing between the $S$ and the $T_i$
fields. This one-loop mixing term with coefficient $\delta_{GS}^i$
generalizes the Green--Schwarz mechanism \GS\ and cancels anomalies
of the underlying non-linear $\sigma$-model \DFKZ,\LOUIS,\CAOV,
which are described
by triangle diagrams with two external gauge bosons and several external
moduli fields $T_i$.
As we will see in the following, the function
$Y=S+\bar S-{1\over 8\pi^2}\sum_{i=1}^3
\delta_{GS}^i\log(T_i+\bar T_i)$ can be regarded
as the redefined string (gauge) coupling
constant at the unifying string scale $M_{\rm string}$: $Y={2\over
g^2_{\rm string}}$. This fact will turn out to be important
for the discussion of the gaugino condensation.

Now, for the one-loop K\"ahler potential to transform in the
required way \kptra\ under target space duality transformations
the dilaton has to acquire a non-trivial modular transformation behaviour
at the one-loop level \DFKZ:
$$S\rightarrow S-{1\over 8\pi^2}\sum_{i=1}^3\delta_{GS}^i
\log(ic_iT_i+d_i).
\eqn\diltra
$$

Next consider the one-loop corrections to the gauge coupling constant
(up to a small field independent contribution):
$$\eqalign{
{1\over g_{\rm 1-loop}^2(\mu)}
&={Y\over 2}+{b_0\over 16\pi^2}
\log{M_{\rm string}^2\over\mu^2}\cr &-
{1\over 16\pi^2}\sum_{i=1}^3({b_0\over 3}
-\delta_{GS}^i)\log\lbrack (T_i+\bar T_i)
|\eta(T_i)|^4\rbrack .\cr}
\eqn\loopgauge
$$
Here $b_0$ is the usual $N=1$ $\beta$-function coefficient,
$b_0=-3C(G)$,
where $C(G)$
is the quadratic Casimir of
the gauge group $G$.
The moduli-dependent last term in eq.\loopgauge\ has
two different sources.
(Up to this term the gauge coupling constant at $\mu=M_{\rm string}$
is given
by $Y/2$, which motivates the previous definition $g_{\rm string}
^{-2}=Y/2$.)
First, at one loop
the massless gauginos lead to the non-harmonic term proportional
to $\sum_{i=1}^3b_0/3\log(T_i+\bar T_i)$. This term is directly
related to the already mentioned $\sigma$-model anomalies
\DFKZ,\LOUIS,\CAOV.
Specifically, the $\sigma$-model anomalies and also the target
space modular anomalies
associated
to a modulus $T_i$ are completely cancelled by the
Green--Schwarz counter term, which follows from eq.\kploop,
if $all$ orbifold twists act non-trivially on the corresponding
$i^{\rm th}$ complex plane of the underlying six-torus.
Then one has $\delta_{GS}^i=b_0/3$. Examples for this situation
are the ${\bf Z}_3$ and ${\bf Z}_7$ orbifolds, where $\delta_{GS}^i=
b_0/3$ for all three complex planes or the ${\bf Z}_4$ orbifolds
with $\delta_{GS}^i=b_0/3$ for two of the three complex planes.

On the other hand
there is in principle no reason that the modular and
$\sigma$-model anomalies have to be completely cancelled
by the Green--Schwarz mechanism. In fact, in general
one has $\delta_{GS}^i\neq b_0/3$ as it is for example true for $T_3$
of the ${\bf Z}_4$ orbifold and also for all three $T_i$ considering
general ${\bf Z}_M\times {\bf Z}_N$ orbifolds. In those cases
there is a second moduli-dependent contribution to the gauge
coupling constant. This contribution describes just the one-loop
threshold
effects \KAPLU\
of the massive string excitations (momentum and winding states)
and is proportional to
$\sum(b_0/3-\delta_{GS}^i)\log |\eta(T_i)|^4$ \DKLB,\FILQ,\DFKZ,\ANT,
where $\eta(T)$ is the well-known Dedekind function.
For the case of complete Green--Schwarz anomaly cancellation,
the contribution from the massive states is absent since the relevant
massive spectrum is then organized into $N=4$ supermultiplets.
Now the remaining target space modular anomaly is exactly removed
by the threshold contributions of the massive states, i.e. the
expression \loopgauge\ is explicitly target space modular-invariant.

Taking into account only the threshold piece of the massive
states,
one recognizes that the one-loop  gauge coupling constant is given
by the real part of a holomorphic gauge kinetic function
of the following form:
$$f_{\rm 1-loop}=S-{1\over 8\pi^2}\sum_{i=1}^3({b_0
\over 3}-\delta_{GS}^i)
\log\eta(T_i)^2.\eqn\gaugekin
$$
It is important to stress that this gauge kinetic function
does not get further renormalized beyond one loop and that it is
therefore
an exact expression at all orders \NIL,\ANT.

For some purposes it will turn out to be convenient to
perform a holomorphic field redefinition
to use a target space modular-invariant dilaton
field $S'$ defined as follows \DFKZ:
$$S'=S+{1\over 8\pi^2}\sum_{i=1}^3\delta_{GS}^i
\log\eta(T_i)^2.\eqn\sredef
$$
Then the string coupling constant $Y$ looks like
$Y=S'+\bar S'-{1\over 8\pi^2}\sum_{i=1}^3
\delta_{GS}^i\log\lbrack(T_i+\bar T_i)|\eta(T_i)|^4\rbrack$,
and the holomorphic one-loop gauge kinetic function takes the
form $f_{\rm 1-loop}=S'-{1\over 8\pi^2}\sum_{i=1}^3{b_0\over 3}
\log\eta(T_i)^2$.

Now let us apply the above results and analyse the modular-invariant
gaugino condensation mechanism first in a hidden
sector, which is described by a
pure Yang--Mills gauge theory with gauge group $G$.
Examples of pure gauge hidden sectors can be found in the literature.
For instance, the $E_8'$ gauge group for the standard embedding
of ${\bf Z}_M$ and ${\bf Z}_M\times {\bf Z}_N$ orbifolds
is most well known. Also ${\bf Z}_7$,
${\bf Z}_4$ models with $SU(3)\times SU(2)\times U(1)^5$
observable sector and pure $E_6$, $SU(3)\times SU(3)$ gauge
hidden sectors respectively were constructed \ORBICON.
There are two consistent approaches
for a dynamical description of the duality-invariant
condensation of gauginos, namely
the effective Lagrangian approach \FMTV,
which contains the gauge-singlet gaugino
bound state as a dynamical degree of freedom, and the
effective superpotential
approach \FILQ\ where the gaugino condensate has been replaced by its
vacuum expectation value.
Both approaches have been shown to be equivalent in refs.\CLMR,\LT,\CM,
and
the effective non-perturbative superpotential is derived as
$$W_{\rm np}\sim
e^{{24\pi^2\over b_0}f}.\eqn\suppotf
$$
Using eq.\gaugekin, the effective superpotential for the case
of a pure Yang--Mills gauge group has the form
$$W_{\rm np}
\sim{e^{{24\pi^2\over b_0}S}\over \prod_{i=1}^3\lbrack\eta(T_i)\rbrack^
{(2-{6\over b_0}\delta_{GS}^i)}}.\eqn\suppotfin
$$
It is important to stress that the effective non-perturbative
superpotential entirely follows from the holomorphic gauge
kinetic function whose moduli dependence originates from
the threshold effects of the massive states, whereas the non-harmonic
contribution
of the massless field does not enter the holomorphic superpotential.
Since the gauge kinetic function does not get renormalized
beyond one-loop, one has obtained an all order expression for the
effective superpotential. It is also important to remark that the
superpotential \suppotfin\ has exactly the correct modular transformation
behaviour \suppottra\ due to the transformation rules of the Dedekind
function and the dilaton field eq.\diltra.
This observation provides very
strong confidence in the non-perturbative validity of target space
modular-invariance.
The correct modular transformation behaviour becomes even more
transparent when using the modular-invariant dilaton field $S'$
and the corresponding gauge kinetic function $f$. Then
the superpotential has an universal moduli dependence,
$$W_{\rm np}\sim{e^{{24\pi^2\over b_0}S'}\over \prod_{i=1}^3\eta(T_i)^2},
\eqn\suppotfina
$$
which is exactly of the form discussed in refs.\FLST\ and \FILQ\ upon
identification of three moduli $T_i$. Of course, the physical
implications will not depend on which version of the non-perturbative
superpotential is used to describe the gaugino condensate, as we will
show in the following.

For analysing supersymmetry breaking
it is necessary
to determine the scalar potential of the theory. The one-loop
contribution to the K\"ahler function in eq.\kploop\ implies
that the formula for the scalar potential studied up to now \FILQ\
must be modified. In particular, taking into account the new K\"ahler
metric, which leads to mixing between the dilaton and the moduli,
one obtains for the scalar potential
$$\eqalign{V={1\over Y\prod_{i=1}^3(T_i+\bar T_i)}&\biggl\lbrace
|W-YW_S|^2+\sum_{i=1}^3{Y\over Y-{1\over 8\pi^2}\delta_{GS}^i}
|(W-{1\over 8\pi^2}\delta_{GS}^iW_S)\cr
&-(T_i+\bar T_i)W_{T_i}|^2-3|W|^2
\biggr\rbrace,\cr}\eqn\scapot
$$
where $W_{\phi}=\partial W/\partial\phi$, $\phi=S,T_i$. The above formula
can be applied for any superpotential. For the case of just one gaugino
condensate in a pure
Yang--Mills hidden sector, the superpotential eq.\suppotfin\ leads to
the following scalar potential:
$$\eqalign{V_{\rm np}
=&{e^{{24\pi^2\over b_0}Y}\over Y\prod_{i=1}^3|\eta(T_i)^2
(T_i+\bar T_i)^{1/2}|^{(2-{6\over b_0}\delta_{GS}^i)}}\biggl\lbrace
|1-{24\pi^2\over b_0}
Y|^2\cr &+\sum_{i=1}^3{Y\over Y-{1\over 8\pi^2}\delta
_{GS}^i}(2-{6\over b_0}\delta_{GS}^i)^2{(T_i+\bar T_i)^2\over 16\pi^2}
|\hat G_2(T_i)|^2-3\biggr\rbrace,\cr}\eqn\scapotfin
$$
where $\hat G_2(T_i)=-{2\pi\over T_i+\bar T_i}-{4\pi\over\eta(T_i)}
{\partial\eta(T_i)\over\partial T_i}$.

This result has various interesting aspects to be discussed.
Using the variables $Y$ and $T$, the scalar potential
looks very similar to the potential in \FILQ, which was derived
without taking into account the loop modifications of the K\"ahler
potential. (The potential of \FILQ\ is rederived for $\delta_{GS}^i=0$
and $T=T_1=T_2=T_3$.)
Moreover, if $\delta_{GS}^i\neq b_0/3$ this potential exhibits
a minimum at $T_i= O(1)$, thus dynamically determining the size of the
$i^{\rm th}$ two-dimensional complex plane.
On the other hand it is easy to realize
that if some moduli do not appear in the superpotential
\suppotfin\ (i.e. $\delta_{GS}^i=b_0/3$), which is the case for
${\bf Z}_M$ orbifolds,
then the scalar potential
is flat with respect to $T_i$. (This non-surprising
behaviour is not automatic
but it arises only after defining the variables $Y$ and $T$, since
there is a non-trivial $T_i$-dependence in the K\"ahler potential.)
Thus for the ${\bf Z}_3$ and ${\bf Z}_7$ orbifolds the scalar potential
possesses no moduli-dependence at all.
The flatness of the potential with respect to the moduli of completely
rotated planes opens
in principle the
possibility of having a large internal radius as discussed in ref. \ANTO.
However there is still the alternative mechanism for the
final goal that all the moduli dynamically acquire vacuum expectation
values, determining the size of the compactified space,
through the non-trivial $T_i$ dependence of the
(twisted) Yukawa couplings as pointed out in refs. \GRA\ and \CLMR.
As it was shown
in ref.\CGM, only if orbifold twists (involved in a particular coupling)
act non-trivially on the $i^{\rm th}$
complex plane, the moduli $T_i$ appear in the final expression
of the Yukawa coupling. These are precisely the moduli that never appear
in the string threshold corrections and hence in the non-perturbative
superpotential obtained from the gaugino condensation. So there is a
kind of completeness in the role of the different moduli in the
perturbative and non-perturbative parts of the superpotential.

Also note that although the ``dilaton'' $Y$ appears in the second term
of eq.\scapotfin, the scalar potential unfortunately exhibits no
minimum with respect to $Y$. Therefore the situation is not improved
concerning the run-away dilaton behaviour after the inclusion
of the loop effects into the K\"ahler potential, and additional
dynamics, as for example several gaugino condensates \TWOGAU,\CLMR,\CM,
\CMA\ or $S$-field duality \SDUAL,
is needed to overcome this serious obstacle.
As already indicated, the use of the superpotential \suppotfina\ with
the universal $T_i$ dependence does not alter at all the
above conclusions.
Now the K\"ahler potential contains the field $S'$ and some additional
moduli dependence. The corresponding scalar potential becomes
$$\eqalign{V&={1\over Y\prod_{i=1}^3(T_i+\bar T_i)}\biggl\lbrace
|W-YW_{S'}|^2-3|W|^2
+\sum_{i=1}^3{Y\over Y-{1\over 8\pi^2}\delta_{GS}^i}\cr &
|W-{1\over 8\pi^2}\delta_{GS}^iW_{S'}(1+2(T_i+\bar T_i){\partial\eta
(T_i)/\partial T_i
\over\eta(T_i)})
-(T_i+\bar T_i)W_{T_i}|^2
\biggr\rbrace.\cr}\eqn\scapota
$$
Using the superpotential \suppotfina\ one ends up again exactly
with the scalar potential \scapotfin.

Let us now turn to the case with massive hidden matter \LT,\CM,\KLN.
In fact, it is known that the existence of hidden matter is the
most general situation and occurs in all promising string
constructions \ORBIM,\FLIP.\foot{It is worth noticing \CLMR,\CM,\CMA\
that in order to fix
the dilaton to a realistic value by using the two gaugino
condensation mechanism, the existence of hidden matter
is crucial.}
For orbifold
compactifications, several explicit condensing
examples can be found. For instance,
in ref.\ORBIM\  phenomenologically interesting
${\bf Z}_3$ models with $G=SU(3)\times SU(2)\times U(1)_Y
\times\lbrack SO(10)\rbrack '$ and three ${\underline{16}}$'s hidden matter
representations was constructed. In ref.\ORBIMA\ two ${\bf Z}_7$ models
with the $SU(3)\times SU(2)\times U(1)^5$ observable gauge group and
$SO(10)$,
$SU(5)\times SU(3)$ hidden sectors with
hidden matter ${\underline{10}}$, $3(\underline 5+
{\bar{\underline 5}})
+7(\underline 3+{\bar{\underline 3}})$
respectively were constructed.
Also a ${\bf Z}_4$ model with $E_6\times SU(2)$
observable sector and
$E_7$ hidden gauge group plus $2({\underline{56}})$  hidden matter
can be found in ref. \KATSUKI.

Specifically, we consider gauge non-singlet
chiral hidden matter fields $Q_\beta$, which become massive through
the coupling to gauge singlet chiral fields $A_\alpha$. These fields
have in any
order of string perturbation theory a completely flat potential
such that $\langle A_\alpha\rangle$ is a perturbatively undetermined
parameter of the theory. Examples for this kind of flat
directions are Wilson line moduli. Thus we consider
the following trilinear perturbative superpotential:
$W=h_{\alpha\beta\gamma}(T_i)A_\alpha Q_\beta Q_\gamma$.
The fields $A_\alpha$, $Q_\beta$ and $Q_\gamma$ have
modular weights $n_{A_\alpha}^i$,
$n_{Q_\beta}^i$, $n_{Q_\gamma}^i$, respectively.
Thus the intermediate
masses for the fields $Q_\beta$ and $Q_\gamma$ are given
by $M_I=|\partial^2 W/(\partial Q_\beta\partial Q_\gamma)|
M_{\rm string}=|h_{\alpha\beta\gamma}(T_i)A_\alpha |
M_{\rm string}$.\foot{The physical
mass of the normalized fields has some additional, non-analytic,
$T_i$ dependence.}
Remember that $h_{\alpha\beta\gamma}$ is moduli dependent
if all fields $A_\alpha$, $Q_\beta$ and $Q_\gamma$ are in the
twisted sector of the orbifold. Otherwise it is constant.

It also happens in several cases that the
fields $A$ give masses $M_I$  to
some vector fields $V$ and additional chiral fields $Q'$
through $D$-term couplings. (The fields $Q'$ build the
longitudinal components of the massive vector bosons $V$.)
Then the vector fields $V$ become
massless for $A=0$, leading to an enlargement of the gauge group.
The breaking of the enlarged gauge group $G'$ down to $G$ by the
vacuum expectation values of the moduli fields
is the so-called stringy Higgs effect \HIGGS.
An especially interesting situation arises when considering orbifolds
with the $i^{\rm th}$ complex plane rotated by all orbifold twists
and with an associated untwisted modulus $A_i$. Then one can show
that, analogous to the modulus $T_i$, the (untwisted)
massive string spectrum
with $A_i$-dependent masses is $N=4$ supersymmetric.
Specifically, the fields $Q$ and
$Q'$ are also in the untwisted sector of the orbifold with
$\underline R_Q=\underline R_{Q'}=\underline R_V$,
and there are twice as many chiral fields $Q$ as fields $Q'$, i.e.
$N_Q=2N_{Q'}=2N_V$.
(The modular weights are for example distributed
in the following way: $n^i_{A_1}=n^i_{Q'}=(-1,0,0)$, $n^i_{Q_2}=
(0,-1,0)$, $n^i_{Q_3}=(0,0,-1)$.)
Then
the fields $Q$, $Q'$ and $V$ build perfect $N=4$ vector supermultiplets.
It follows that in this case
the ``mass'' field $A$ does not enter the non-perturbative superpotential
as we will discuss now.

Let us analyse the case of intermediate masses higher than
the condensation scale $\Lambda$, i.e. $\Lambda <M_I
 <<
M_{\rm string}$. The one-loop hidden gauge coupling constant
is of the form
$$\eqalign{
{1\over g_{\rm 1-loop}^2(\mu)}
&={Y\over 2}+{b_0\over 16\pi^2}
\log{|h_{\alpha\beta\gamma}(T_i)A_\alpha |^2M^2_{\rm string}
\over\mu^2}-{b_1\over 16\pi^2}\log
|h_{\alpha\beta\gamma}(T_i)A_\alpha |^2
\cr -&
{1\over 16\pi^2}\sum_{i=1}^3(b_0/3
-\delta_{GS}^i)\log (T_i+\bar T_i)
-{1\over 16\pi^2}\sum_{i=1}^3(b_i'
-\delta_{GS}^i)\log
|\eta(T_i)|^4 .\cr}
\eqn\loopgaugem
$$
Here $b_0=-3C(G)$ is the $N=1$ $\beta$-function coefficient where
all matter fields are decoupled. On the other hand,
$b_1$ describes the running between $M_I$ and $M_{\rm string}$
where all fields contribute and is therefore given by
$b_1=-3C(G)-3\sum_V T(\underline R_V)
+\sum_Q T(\underline R_Q)
+\sum_{Q'} T(\underline R_{Q'})$.
The threshold corrections\foot{Strictly
speaking, the mass of the momentum and winding states should not only
depend on the moduli $T_i$ but also on the field
$A$. Therefore
the Dedekind function in eq.\loopgaugem\  provides an expression for
the determinant of the heavy field mass matrix, which is only valid for
$M_I<<M_{\rm string}$. For arbitrary values of $A$,
$\eta(T_i)$ should
be replaced by a more general automorphic function,
which involves both $T_i$ and $A$
but still transforms under target modular transformations like $\eta
(T_i)$.}
due to the massive momentum and winding states are determined
by the
coefficient $b_i'$, where all massless gauginos of $G$ as well as
the massive fields $Q$ and possibly $Q'$, $V$ ``contribute''
in the following way \LOUIS,\DFKZ: $b_i'=-C(G)-\sum_V T(\underline R_V)
+\sum_Q T(\underline R_Q)(1+2n_Q^i)
+\sum_{Q'} T(\underline R_{Q'})(1+2n_{Q'}^i)$.
Note that the $\sigma$-model anomalies, related
to $\log(T_i+\bar T_i)$, are only generated by the contribution
of the massless gauginos of $G$. However the coupling
constant $g(\mu)^{-2}$ is still target space modular-invariant
due to the non-trivial transformation property of $h_{\alpha\beta\gamma}
(T_i)A_\alpha$.
This follows immediately if there are no states $Q'$ and $V$.
Otherwise the requirement of target space modular-invariance
imposes the following restriction on the additional states:
$$\sum_VT(\underline R_V)(-2-3n_{Q_\beta}^i-3n_{Q_\gamma}^i)+
\sum_{Q'}T(\underline R_{Q'})(n_{Q_\beta}^i+n_{Q_\gamma}^i-2n_{Q'}^i)=0.
\eqn\restr
$$
The fact that the massive matter fields
have the same effect as the massless matter fields with respect
to modular transformations was already noted in ref.\IBLU\
in the context of strong $CP$ violation.

Consider also the special case of a completely rotated plane $i$
with an associated untwisted modulus $A_i$. Then
the chiral fields
$Q$ and $Q'$, being also untwisted,
build together with the vectors $V$
$N=4$ vector supermultiplets, and one therefore obtains $b_1=b_0$.
This implies that the running
coupling constant does not depend on $A_i$.
In the same way it follows that
$b_i'=-C(G)=b_0/3$.
This can be explicitly checked in several examples \IBPR.
Thus for planes which are rotated by all orbifold twists
we still have  $b_i'=b_0/3=\delta_{GS}^i$ as in the pure
Yang--Mills case.

The gauge kinetic function, including the holomorphic contribution of the
massive states to the gauge coupling constant, takes the following form:
$$f_{\rm 1-loop}=S+{b_0-b_1\over 8\pi^2}\log(h_{\alpha\beta\gamma}(T_i)
A_\alpha)-{1\over 8\pi^2}\sum_{i=1}^3(b_i'-\delta_{GS}^i)\log\eta(T_i)^2.
\eqn\gaugekinfm
$$
Using the same arguments as
in the pure Yang--Mills case, the non-perturbative
superpotential is given by $W_{\rm np}\sim
e^{{24\pi^2\over b_0}f}$. Then we
obtain with eq.\gaugekinfm\foot{For the case $M_I<\Lambda$ one
obtains the same form for the non-perturbative superpotential
including $Q_\beta Q_\gamma$ bound states into the effective
Lagrangian.}
$$W_{\rm np}\sim{e^{{24\pi^2\over b_0}S}\lbrack
h_{\alpha\beta\gamma}(T_i)A_\alpha\rbrack^{3(b_0-b_1)
/b_0}
\over \prod_{i=1}^3\lbrack\eta(T_i)\rbrack^
{6(b_i'-\delta_{GS}^i)/b_0}}.\eqn\suppotfinm
$$
The
above superpotential has the correct modular transformation behaviour
\suppottra\ due to the transformation rules of the Dedekind function,
dilaton and matter fields eqs.\diltra,\mattrans\ and Yukawa couplings
eq.\yuktrans.
It is important to stress the fact that when the matter representations
acquire mass through twisted Yukawa couplings (i.e. $h_{\alpha\beta\gamma}
\neq{\rm const}$),
the whole set of moduli appears in the gaugino condensation
superpotential \suppotfinm. Therefore even for moduli $T_i$ with
the $i^{\rm th}$ plane rotated by all orbifold twists,
i.e. $b_i'=\delta_{GS}^i$,
there will be no flat directions.
On the other hand, for the particular case of
a completely rotated plane with an associated
untwisted field $A_i$,
both  $T_i$ and  $A_i$ do not appear in $W_{\rm np}$
since $b_0=b_1=3b_i'=3\delta_{GS}^i$ as discussed before.

In order to analyse supersymmetry breaking, one has to include
in the K\"ahler potential the contribution of the matter fields.
Let us study the case of untwisted matter\foot{For an orbifold
example with all the hidden matter
in the untwisted sector, see ref. \KATSUKI.},
$C_i$,
where the string tree level K\"ahler potential is known
at all orders in $C_i$,
$K_{\rm tree}=-\log (S+\bar S)-\sum_{i=1}^3
\log X_i$,
where $X_i=T_i+\bar T_i-|C_i|^2$.
Therefore it is very plausible to assume that the one-loop
modification is given by
$$K_{\rm 1-loop}
=-\log Y-\sum_{i=1}^3\log X_i,
\eqn\kploopm
$$
with $Y=S+\bar S-{1\over 8\pi^2}
\sum_{i=1}^3
\delta_{GS}^i\log X_i$.
$K_{\rm 1-loop}$ leads to mixing between the dilaton, $T_i$ and $C_i$
fields, and one obtains for the scalar potential
$$\eqalign{V={1\over Y\prod_{i=1}^3 X_i}&\biggl\lbrace
|W-YW_S|^2+\sum_{i=1}^3{Y\over Y-{1\over 8\pi^2}\delta_{GS}^i}\lbrack
|(W-{1\over 8\pi^2}\delta_{GS}^iW_S)\cr
&-X_iW_{T_i}|^2+
X_i|W_{C_i}+\bar C_iW_{T_i}|^2\rbrack
-3|W|^2
\biggr\rbrace,\cr}\eqn\scapotm
$$
The above formula can be applied for any superpotential.
Let us study the case of untwisted
matter $Q_j$, $Q_k$ coupled to $A_i$ ($i\neq j\neq k\neq i$).
For the case
of just one gaugino condensate with hidden matter fields, the
superpotential eq.\suppotfinm\ leads to the following scalar potential:
$$\eqalign{V_{\rm np}
=&{e^{{24\pi^2\over b_0}Y}|A_i|^{6(b_0-b_1)/b_0}
\over Y\prod_{l=1}^3|\eta(T_l)|^{12(b_l'-\delta_{GS}^l)/b_0}
X_l^{(1-{3\over b_0}\delta_{GS}^l)}}\biggl\lbrace
|1-{24\pi^2\over b_0}Y|^2-3
\cr &+\sum_{l=1}^3{Y\over Y-{1\over 8\pi^2}\delta
_{GS}^l} |(1-{3\delta_{GS}^l\over b_0})+X_l{6\over b_0}
(b_l'-\delta_{GS}^l)
{\partial\eta(T_l)/\partial T_l\over \eta(T_l)}|^2
\cr &+{Y\over Y-{1\over 8\pi^2}\delta_{GS}^i}{X_i\over |A_i|^2}
|3(1-{b_1\over b_0})-|A_i|^2{6\over b_0}(b_i'-\delta_{GS}^i)
{\partial\eta(T_i)/\partial T_i\over \eta(T_i)}|^2
\biggr\rbrace,\cr}\eqn\scapotfinm
$$
where $X_l=(T_l+\bar T_l)$ for $l\neq i$. As discussed in ref.\LT, by
minimizing this potential with respect to $T_l$ ($j=1,2,3$) and
$A_i$, one generically obtains  that the vacuum expectation values
of all fields $T_l$, $A_i$ are dynamically determined for generic
values of the parameters $b_0$, $b_1$, $b_l'$ and $\delta_{GS}^l$.
The only exception is again provided if the complex plane $i$
is rotated by all orbifold twists. Then $T_i$ and $A_i$
do not appear in the scalar potential and take arbitrary vacuum
expectation values.
Also note that the ``dilaton'' $Y$ still possesses the run-away
behaviour as in the pure Yang--Mills case.

\bigskip
\bigskip
\ack
We would like to thank J.A. Casas and L.E. Ib\'a\~nez for very useful
discussions.

\par \penalty-400 \vskip\chapterskip
   \spacecheck\referenceminspace \immediate\closeout\referencewrite
   \referenceopenfalse
   \line{\fourteenrm\hfil REFERENCES\hfil}\vskip\headskip
   \input referenc.texauxil
   
\vfill\eject\bye